# Polarization rotator Bragg grating assisted wavelength selective polarization alignment

HIDEAKI OKAYAMA,[1,2,*] YOSUKE ONAWA,[1,2] DAISUKE SHIMURA[1,2], HIROYUKI TAKAHASHI[1,2], HIROKI YAEGASHI[1,2] AND HIRONORI SASAKI[1,2]

[1] R&D Center, Oki Electric Industry Co., Ltd., Warabi, Saitama 335-8510, Japan
[2] Photonics Electronics Technology Research Association (PETRA), Warabi, Saitama 335-8510, Japan
*okayama575@oki.com

**Abstract:** A device performing both the polarization-alignment and the wavelength selection is demonstrated by fabricated device. Waveguides incorporating Bragg gratings for polarization rotation and mode conversion having different widths are placed near to each other. For the polarization rotation, experimental results show that the field strength ratio in the two waveguides defines the diffraction coupling coefficient and excitation strength of unwanted modes. The excitation of unwanted mode is reduced, when a large radius short curved waveguide is used for input/output waveguides as shown in the experiment. Polarization aligned wavelength peaks were observed at the backward drop port in the experiment.

## 1. Introduction

The silicon waveguide technology is [1-18] studied extensively. In the optical communication technology, as the light signal transmitted through optical fiber has random polarizations and the waveguide device tends to operate in single polarization, a scheme to control polarization is required. Many experimental polarization rotation Bragg grating devices [19-28] are reported to solve the problem. However, polarization independence is often achieved by polarization alignment used in the polarization diversity scheme. Here, we report a device for this scheme.

Back-diffracted light can be separated using single grating without circulator by a contra-directional coupling of diffracted light in asymmetric directional coupler [12-16]. We showed that by proper design a single polarization rotation grating waveguide coupled to another waveguide can separate the polarization rotated and diffracted light [29]. The device can be used in wavelength selective aligning of the polarizations using combination with TE0/TE1 mode order conversion as will be shown. Firstly, we show the experimental verification of the design and details of the requirement for the input/output waveguide using experimental results obtained after our preliminary device experimental report [30]. We designed the device around 1600 nm for next generation wavelength division multiplexing (WDM) fiber to the home system (FTTH). The device is to be used in polarization diversity as such that the desired wavelength band is selected and the polarization is aligned to the TE mode simultaneously. The unique feature of this device may also find many applications not envisioned at present. We used finite element method (FEM) to calculate normal modes and three-dimensional finite difference time domain (3D-FDTD) simulation to obtain the theoretical design conditions.

In the next section we describe the basic of the polarization rotation. In section 2.2 we show the example of the wavelength response obtained by experiment. In section 2.3 the effect of the optical field strength ratio between two waveguides on diffraction strength is described. The effect of input/waveguide structure is discussed in section 2.4. A device performing both the polarization-alignment and the wavelength selection using this concept is shown in section 3. The paper is summarized in section 4.

## 2. Basic coupler-grating polarization rotator

*2.1 Device structure*

The device structure for converting the TM mode to the TE mode is shown in Fig. 1. Two waveguides with different widths incorporating polarization rotation Bragg grating are used. In Fig. 1 polarization rotation Bragg grating is placed in the narrower waveguide. An antisymmetric grating and the waveguide structure asymmetric in the depth direction is required to generate polarization rotation. In this report, as in Fig. 1(B) showing a waveguide cross section, a rib waveguide structure with restricted slab area is used. The Bragg diffraction between even TM and odd TE normal modes is selected. The even TM mode light is generated by injecting a light into the wide waveguide. This mode is converted into odd TE mode by the Bragg diffraction whose power mainly resides in the narrow waveguide. The polarization rotated light is ejected from the narrow waveguide at the Bragg wavelength. This setting avoids using the weak-confined and high-loss odd TM mode. The grating is not placed in the wide waveguide to avoid strong diffraction between even TM and even TE modes.

We use Si waveguide core thickness of 220 nm which is often the standard for many foundries. We use 70 nm half etch depth grating. The grating corrugation is 150 nm. Figure 2 shows the optical field obtained by finite element method (FEM) for modes used in the main diffraction. A 300 nm gap between waveguides and waveguide widths of 540 and 340 nm were used as an example. The sum of overlap integral between the electric field components of even TM and odd TE mode defines the diffraction coefficient [20]. The field symmetry in the figure shows that the above mentioned grating and waveguide structure can generate non-zero diffraction coefficient similar to the single waveguide type polarization rotation Bragg grating.

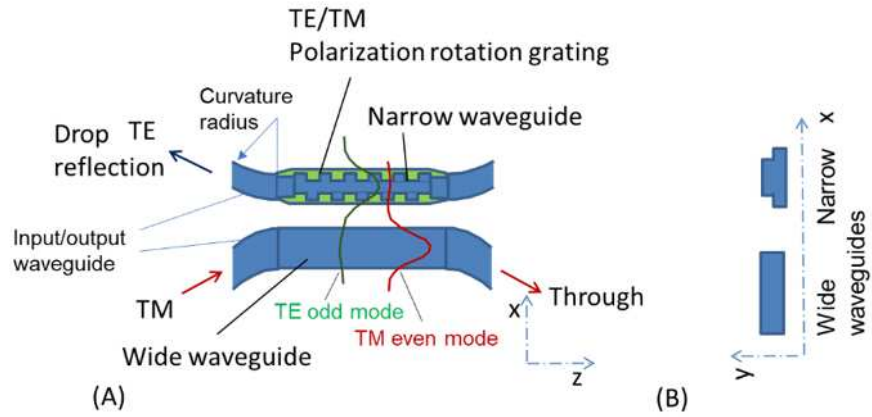

Fig. 1. Asymmetric directional coupler contra-directional polarization rotation Bragg grating structure. (A) device structure and (B) waveguide cross section are shown..

Sub-diffraction at shorter wavelength would be generated when unwanted odd TM mode is excited [29]. The presence of the unwanted odd TM mode results in lower main diffraction, wavelength dip in the through port and crosstalk in the forward drop port. The sub-diffraction wavelength is shorter than that of the main diffraction. We optimized the waveguide gap and the width difference to suppress unwanted transfer of the light between waveguides and to obtain sufficient amount of diffraction as will be explained. We also used 120 μm radius curvature for the input and output waveguides to suppress the mode conversion effect at these waveguides as will be explained.

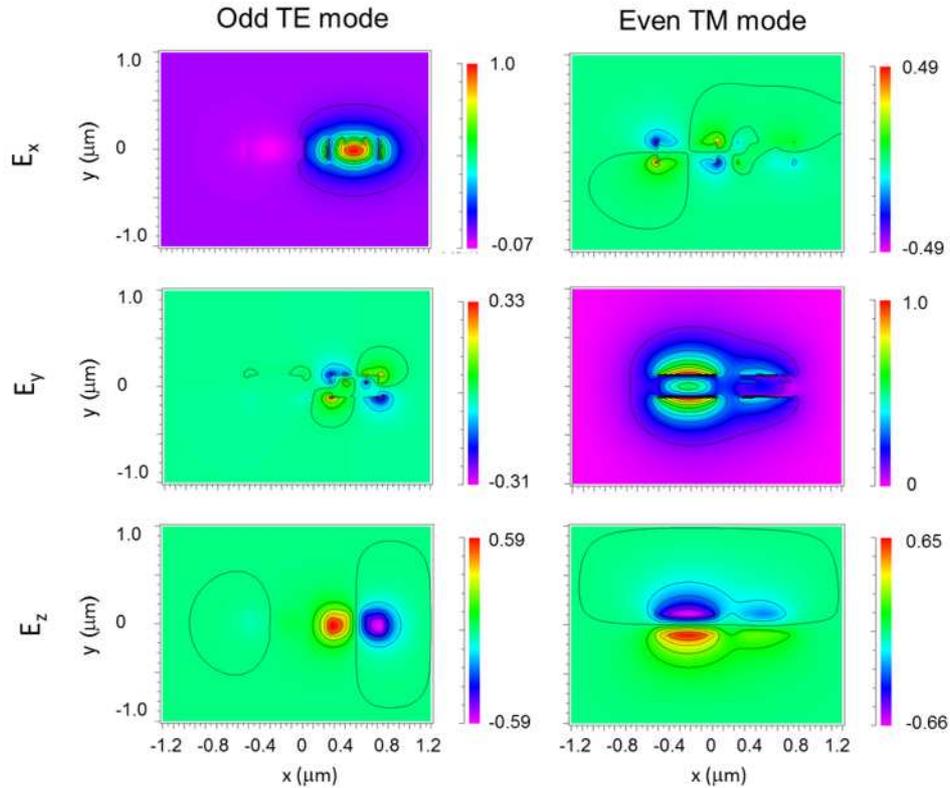

Fig. 2. Calculated optical field in waveguides obtained by FEM. Wavelength is 1600 nm, gap width is 300 nm and 220 nm thick waveguides with 540 and 340 nm widths are used.

## 2.2 Example of experimental wavelength response

The device was fabricated using 300 mm line foundry wafer process. The immersion ArF lithography and reactive ion etching (RIE) were used for defining the waveguide pattern. The 220 nm thick Si waveguide core was covered by $SiO_2$ using chemical vapor deposition (CVD). The waveguide widths are 545 and 375 nm. The gap between wide waveguide and narrow waveguide terrace is 325 nm. The grating period $\Lambda$ is 392 nm. The grating is etched 70 nm from the top surface. Corrugation width is 150 nm. The grating length is 1250$\Lambda$.

Wide wavelength source and optical spectrum analyzer were used to measure the wavelength characteristics of the device. The input light polarization was controlled by polarizer obtaining polarization extinction ratio of 20 dB. The measured results are shown in Fig. 3. A distinct polarization rotated wavelength peak was observed at the backward drop port (p3) when the TM polarization mode was launched into the input port. When the TE mode was launched, no distinct wavelength peak was observed. The wavelength dip at the through port was 22 dB. The insertion loss was 1 dB. The wavelength peak width was 7 nm. A wavelength peak generated by sub-diffraction is observed around 1570 nm wavelength.

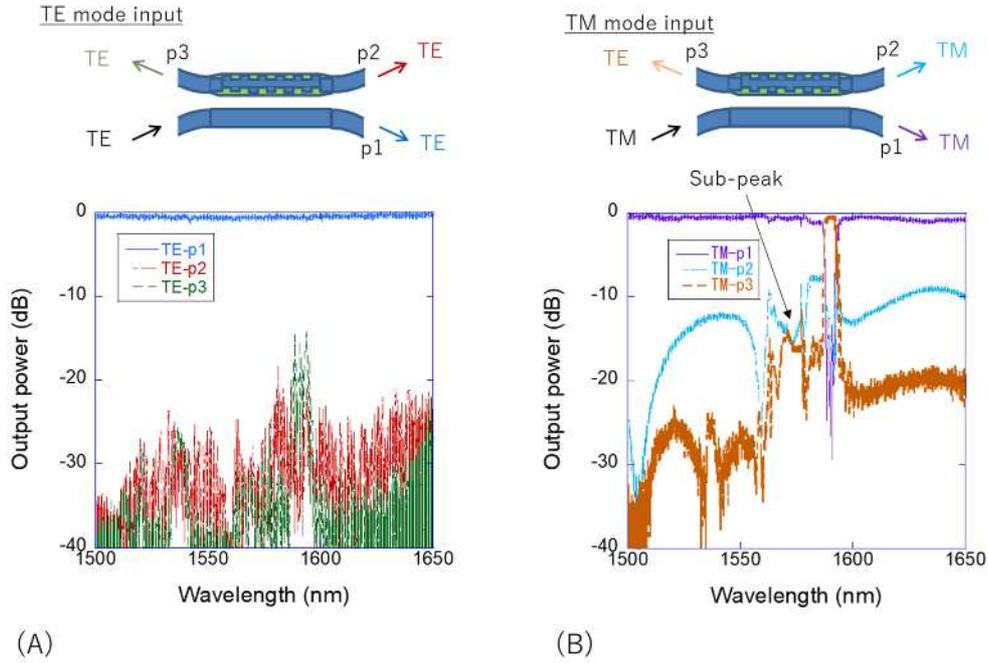

Fig. 3 The experimental wavelength response with 545 and 375 nm waveguide width and 325 nm gap and 1250$\Lambda$ length.

### 2.3 Effect of optical field strength ratio in two waveguides

The ratio of the TM normal mode field strengths in two waveguides $r$ can be used as a guide for the design. The ratio of maximum field strengths is used for simplicity. The odd TE mode optical field in the wide waveguide can be ignored for its smallness as can be noticed in Fig. 2. The diffraction coefficient $K$ can be deduced from the overlap of the even TM and odd TE modes in the narrow grating waveguide, so that the field strength in the grating waveguide defines it. The graph of Fig. 4 shows the diffraction coefficient $K$ between even TM and odd TE modes as a function of field strength ratio $r$ between waveguides obtained by simulation and experiment. The value of $K$ is normalized by fractional corrugation $D=\Delta W/W$, which is ratio of corrugation $\Delta W$ to waveguide-width $W$. The ratio $r$ can be changed by gap width or the waveguide width differences. The ratio $r$ is obtained by FEM. The coupling coefficient is obtained from 3D-FDTD simulation. In Fig. 4 the gap widths of 300, 400, 450 and 500 nm were used with width difference of 200 nm unchanged. The width differences of 100, 140, 200 and 240 nm were used with the gaps of 300, 400 or 500 nm unchanged respectively. The gap is defined between the wide and average narrow grating waveguide edges. Figure 4 shows that the diffraction coefficient $K$ is defined by $r$. A slight discrepancy between the 3D-FDTD results for gap and width difference might be the field shape difference. The experimental diffraction coefficients estimated from the drop port diffraction peak widths and through port dip depths are also shown in Fig. 4 for width differences of 130, 170, 200 and 360 nm (Table 1).

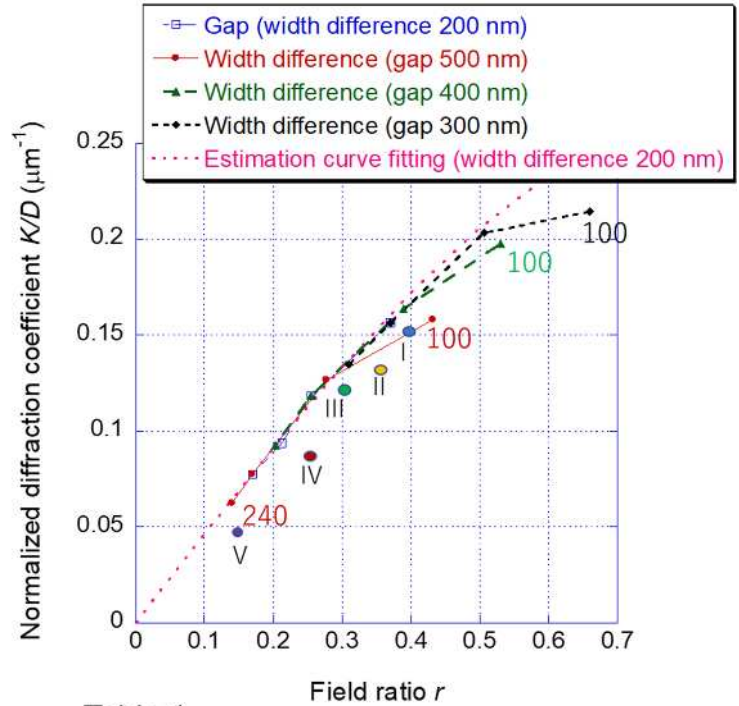

Table 1

| Experimental device No. | Waveguide widths (nm) | | Gap (nm) |
|---|---|---|---|
| I | 525 | 395 | 400 |
| II | 540 | 340 | 300 |
| III | 545 | 375 | 400 |
| IV | 500 | 300 | 400 |
| V | 800 | 440 | 225 |

Fig. 4. Polarization rotation diffraction coefficient K versus ratio of field intensity r between two waveguides. The value of K is normalized by fractional corrugation D, which is ratio of corrugation and waveguide-width. The dependences are obtained by changing the gap width and waveguide width differences. Gap widths of 300, 400, 450 and 500 nm were used with width difference of 200 nm unchanged. Width differences of 100, 140, 200 and 240 nm were used with the gaps of 300, 400 or 500 nm unchanged respectively. The normalized diffraction coefficient estimated from peak width of experiment is shown together.

Larger $r$ value is beneficial in obtaining high diffraction efficiency as shown in Fig. 4. The excitation power rate of the unwanted odd TM mode is $r^2/(1+r^2)$ when a light is directly injected into the wide waveguide abruptly. The excitation of the unwanted odd TM mode results in lower main peak and sub-diffraction at shorter wavelength as described before. Thus smallest possible $r$ is beneficial in this regard but this result in smaller diffraction. Adiabatic excitation of the even TM mode is required so that the excitation of the odd TM mode is suppressed. We found that input and output curved waveguides with large radius can be used to obtain this condition as will be shown. Under optimized design diffraction with enough strength can be obtained with low odd TM mode excitation. We should be reminded that small diffraction coefficient is needed to obtain a narrow transmission peak linewidth which requires some effort in a silicon waveguide type device [9].

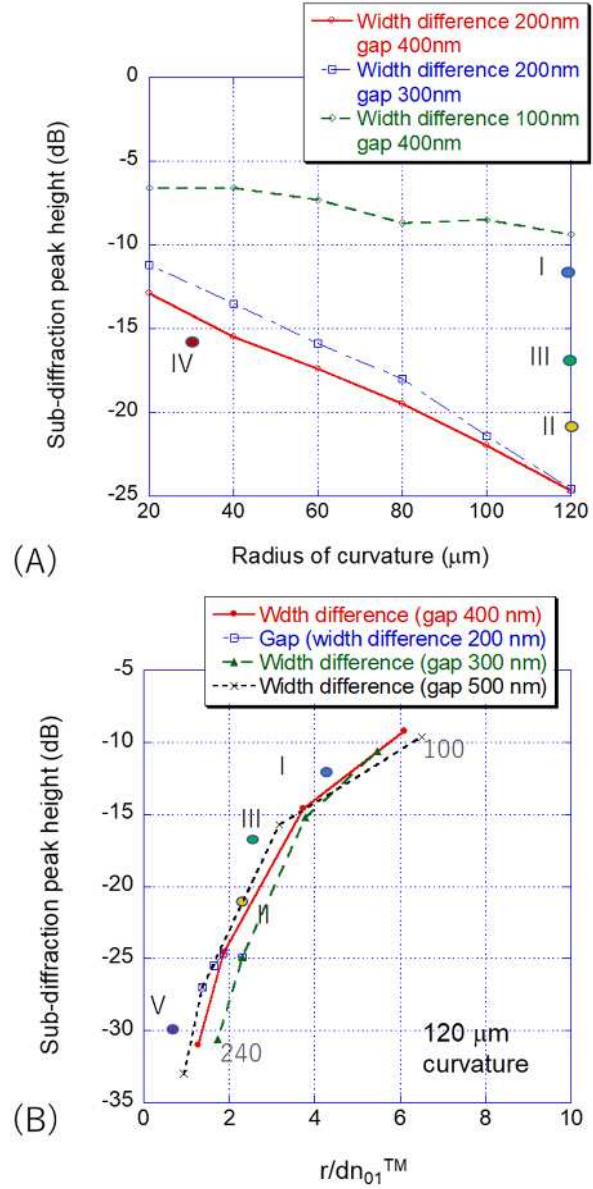

Fig. 5. Height of the unwanted sub-diffraction peak of the odd TM mode as a function of (A) input and output curved waveguide radius and (B) field ratio $r$ normalized by effective index difference between even and odd modes $dn_{01}^{TM}$ obtained by 3D-FDTD simulation. The excitation of the odd TM mode is suppressed by large radius. The field strength ratio of two waveguides is $r = 0.255$ for waveguide gap of 400 nm and 200 nm width difference. For 300 nm waveguide gap and 200 nm width difference $r = 0.370$. For 400 nm waveguide gap and 100 nm width difference $r = 0.530$. Results obtained from experiment are shown together (The design is shown in table of Fig. 4).

*2.4 Effect of input/output waveguides*

A bump on the peak at 1570 nm wavelength in Fig. 3 is the diffraction of unwanted odd TM to odd TE mode. The sub-diffraction bump height as a function of input and output waveguide radius of curvature is shown in Fig. 5(A). The sub-diffraction bump height as a function of field ratio r normalized by effective index difference $dn_{01}{}^{TM}$ between even and odd modes is shown in Fig. 5(B). The results obtained by simulation and experiment for devices listed in Table 1 are shown in Fig. 5. The $dn_{01}{}^{TM}$ is known to affect the conversion between modes [31]. The experimental sub-diffraction peak heights show fair agreement with those obtained by simulation. The bump height decreases for larger radius showing suppression of the odd TM mode excitation. We found that the smallness of $r$ allows easier achievement of adiabatic like condition.

The length of the curved waveguide section is given by $L_r=(2SR-S^2)^{1/2}$ with $R$ being the curvature radius and $S$ being the maximum gap enlargement from the curved waveguide start to the end. For $S=1\mu m$, which is wide enough to prevent coupling, the curved waveguide length is $L_r=15$ μm for $R=120$ μm as those used in the simulation. Thus, the large curvature radius doesn't contribute much to the total length of the device in which the grating section is several hundreds of microns long.

## 3. Combination with TE0/TE1 grating

*3.1 Application to the polarization insensitive optical circuit*

The polarization independent optical circuits composed of elements working only for the TE mode using polarization diversity scheme is shown in Fig. 6. Using elements such as multiple AWG can increase the wavelength channels $\lambda_1$, $\lambda_2$, $\lambda_3$ [32]. We used separate polarization splitter, polarization-aligner and wavelength selective element together with AWGs. The simpler configuration than in Ref. 32 shown in Fig. 6 can be constructed using the device proposed in this section. A single device can perform both wavelength-selection and polarization-alignment functions.

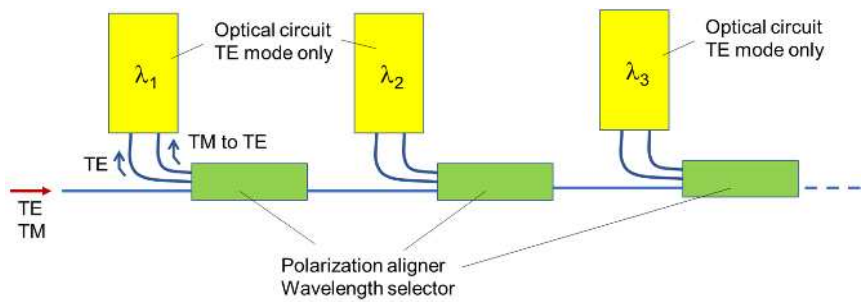

Fig. 6 Configurations of polarization independent optical circuit using diversity scheme.

*3.2 Device combined with TE0/TE1 mode conversion grating*

The device structure is shown in Fig. 7. Among several conceivable configurations [30] for this target, the device we report here showed the best experimental performance so far. Two waveguides with different widths placed in proximity to each other are used. The polarization rotation Bragg grating is formed in the narrow waveguide. The TE0 to TE1 mode conversion Bragg grating [10, 11] is placed in the wide waveguide. A narrow-tapered waveguide is placed near the wide waveguide in front of the mode conversion grating to select and couple-out the diffracted TE1 mode. An antisymmetric grating is used for the TE0 to TE1 mode conversion Bragg grating. The grating formed in a sidewall of the waveguide is shifted half period against the grating of another side [7]. In this grating the simple full-etched structure is used which can maximize the diffraction efficiency. The grating period of $\Lambda=\lambda_B/(N_{TE0}+N_{TE1})$ is required for diffraction at wavelength of $\lambda_B$, where $N_{TE0}$ and $N_{TE1}$ are effective indices of TE0 and TE1 modes in the wide waveguide, respectively.

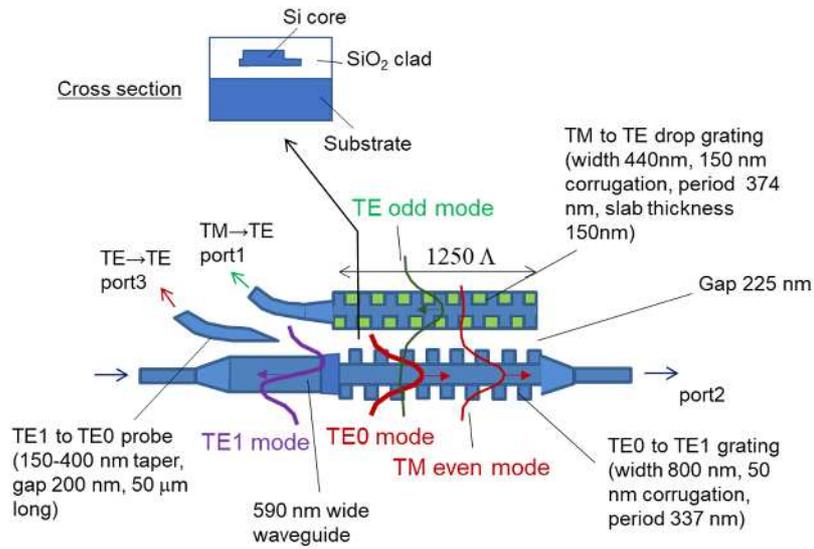

Fig. 7 Asymmetric directional coupler contra-directional polarization rotation Bragg grating structure combined with TE0 to TE1 Bragg grating.

For the polarization rotation grating, the waveguide structure asymmetric in the depth direction and an antisymmetric grating are required [20]. A rib waveguide structure with small slab area is used as in Fig. 7 showing the waveguide cross section. The even TM mode light is excited by injecting a light into the wide waveguide. The Bragg grating generates diffraction between even TM and odd TE normal modes. The odd TE mode power mainly resides in the narrow waveguide. At the Bragg wavelength, the polarization rotated light is ejected from the narrow waveguide. The grating period of $\Lambda=\lambda_B/(N_{TMe}+N_{TEo})$ is used for Bragg wavelength of $\lambda_B$, where $N_{TMe}$ and $N_{TEo}$ are effective indices of even TM and odd TE modes, respectively.

We use Si waveguide core thickness of 220 nm and half-etch depth of 70 nm which are often the standard for many foundries. The widths of 440 nm and 800 nm are used for the narrow and wide waveguides. This width for the narrow waveguide is selected to assure immunity to the width error and obtain sufficient overlap of the modes between those of the narrow and wide waveguides to attain high diffraction efficiency.

The width of the wide waveguide is selected to move the radiative diffraction tend to be seen in this kind of device toward shorter wavelength. This grating is symmetrical in the depth direction so that it doesn't affect the TM to the TE mode diffraction both in the wide waveguide and between narrow and wide waveguides. The asymmetric grating doesn't generate the diffraction between the even and odd modes of the coupler with the same polarization. Thus, unwanted diffraction is minimized in this structure.

The grating period and corrugation are 374 nm and 150 nm for the polarization rotation Bragg grating, respectively. The grating period and corrugation are 337 nm and 50 nm for the TE0 to TE1 mode conversion Bragg grating, respectively. The grating periods must be tuned depending on foundry process conditions. We used raised-cosine type apodized grating structure for both gratings to lower the sidelobe height. The length of each grating is 1250Λ which is around 470 μm. The gap between the narrow and wide waveguides is 225 nm. The waveguide coupler section for coupling-out the TE1 mode to the output port is 50 μm long. The width of the wider waveguide is 590 nm and the narrower waveguide width is tapered from 150 to 400 nm.

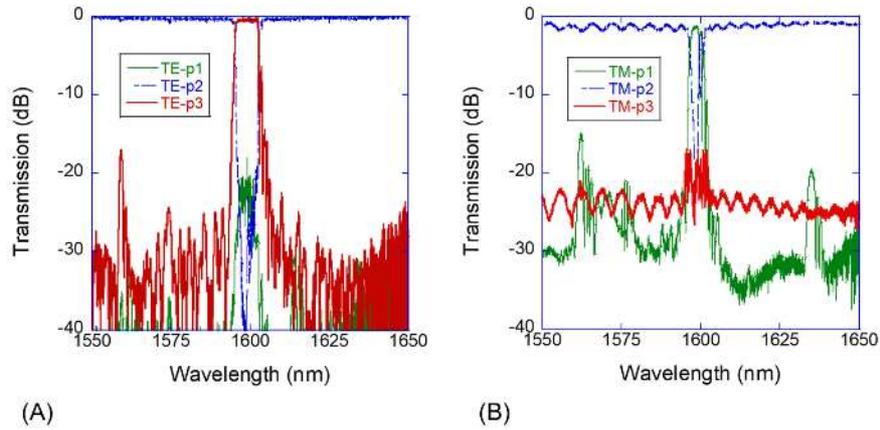

Fig. 8 Measured wavelength response of device shown in Fig. 7 for the (a) TE and (b) TM input lights.

Figure 8 shows the measured wavelength response obtained by preliminary experiment. The device is fabricated in the silicon-on-insulator (SOI) 300 mm line foundry using ArF excimer immersion lithography followed by reactive ion dry etching and $SiO_2$ clad covering using chemical vapor deposition.

A wavelength response expected from the design is obtained. Expected TM to TE diffraction wavelength peak is observed in the drop port of the narrow waveguide (port1). The TE0 to TE1 diffracted light wavelength peak is observed in the output port for the TE mode (port3). We obtained 18 dB and 22 dB extinction ratios for the TM and TE modes, respectively. The crosstalk floor is 22 dB. The excess loss which is the device insertion loss excluding fiber-waveguide coupling and propagation losses is less than 1.5 dB. The wavelength peak widths are 3 nm and 8 nm for the TM and TE modes, respectively. By replacing polarization rotation grating with the TM mode grating and connecting port3 to the input of the waveguide to the port1, we can also obtain a polarization independent wavelength filter.

4. **Summary**

We have reported experimental results of silicon waveguide contra-directional coupler type polarization rotation Bragg gratings for polarization alignment. Experimental results show that the field strength ratio in two waveguides defines the diffraction coupling coefficient and excitation of unwanted modes as indicated from 3D-FDTD simulation. The excitation of unwanted mode is reduced, when a large radius short curved waveguide as input/output waveguides are used as shown in the experiment. A polarization rotated wavelength peak was observed at the backward drop port in the fabricated device. A device performing both the polarization-alignment and the wavelength selection using this concept is proposed and demonstrated by fabricated device.

**Funding.** This paper is based on results obtained from a project, JPNP13004, commissioned by the New Energy and Industrial Technology Development Organization (NEDO).

**Disclosures**. The authors declare no conflicts of interest.

## References

1. T. Tsuchizawa, K. Yamada, H. Fukuda, T. Watanabe, J. Takahashi, M. Takahashi, T. Shoji, E. Tamechika, S. Itabashi, and H. Morita, "Microphotonics devices based on silicon microfabrication technology," IEEE J. Select. Topics Quantum Electron., vol. 11, no. 1, pp. 232-239, Jan/Feb, 2005.
2. H. Yamada, T. Chu, S. Ishida, and Y. Arakawa, "Silicon photonic wire waveguide devices," IEICE Trans. Electron., vol. E90-C, no. 1., pp. 59-64, Jan, 2007.
3. W. Bogaerts, S. K. Selvaraja, P. Dumon, J. Brouckaert, K. De Vos, D. Van Thourhout, and R. Baets, "Silicon-on-insulator spectral filters fabricated with CMOS technology," J. Select. Areas Quantum Electron., vol. 16, no. 1, pp. 33-44, Jan/Feb, 2010.
4. H. Okayama, K. Kotani, Y. Maeno, D. Shimura, H. Yaegashi, and Y. Ogawa, "Design of polarization-Independent Si-wire-waveguide wavelength demultiplexer for optical network unit," Jpn. J. Appl. Phys., vol. 49, pp. 04DG19-1-5, Apr, 2010.
5. C. Cremer, G. Heise, R. Marz, M. Schienle, G. Schulte-Roth, and H. Unzeitig, "Bragg gratings on InGaAsP/InP waveguides as polarization independent optical filters," J. Lightwave Technol., vol. 17, no. 11, pp. 1641-1645, 1989.
6. T. E. Murphy, J. T. Hastings, and H. I. Smith, "Fabrication and characterization of narrow-band Bragg-reflection filters in silicon-on-insulator ridge waveguides," J. Lightwave Technol., vol. 19, no. 12, pp. 1938–1942, 2001.
7. D. T. H. Tan, K. Ikeda, and Y. Fainman, "Cladding-modulated Bragg gratings in silicon waveguides," Opt. Lett., vol. 34, no. 9, pp. 1357-1359, 2009.
8. X. Wang, W. Shi, R. Vafaei, A. F. Jaeger, and L. Chrostowski, "Uniform and sampled Bragg gratings in SOI strip waveguides with sidewall corrugations," IEEE Photon. Technol. Lett., vol. 23, no. 5, pp. 290-292, 2011.
9. X. Wang, W. Shi, H. Yun, S. Grist, N. A. F. Jaeger, and L. Chrostowski, "Narrow-band waveguide Bragg gratings on SOI wafers with CMOS-compatible fabrication process," Opt. Express, vol. 20, no. 14, pp. 15547–15558, 2012.
10. Y. Onawa, H. Okayama, D. Shimura, S. Miyamura, H. Yaegashi and H. Sasaki, "Polarization-insensitive Si wire waveguide add/drop wavelength filter using reflective mode conversion grating and mode split coupler," Electron. Lett. 48(20), 1297-1298 (2012).
11. H. Qiu, J. Jiang, T. Hu, P. Yu, J. Yang, X. Jiang, and H. Yu "Silicon add-drop filter based on multimode Bragg sidewall gratings and adiabatic couplers, " J. Lightwave Technol., vol. 35, no. 9, pp. 1705-1709, 2017.
12. W. Shi, X. Wang, W. Zhang, L. Chrostowski, and N. A. F. Jaeger, "Contradirectional couplers in silicon-on-insulator rib waveguides," Opt. Lett., vol. 36, no. 20, pp. 3999 -4001, 2011.
13. W. Shi, H. Yun, C. Lin, M. Greenberg, X. Wang, Y. Wang, S. T. Fard, J. Flueckiger, N. A. F. Jaeger, and L. Chrostowski, "Ultra-compact, flat-top demultiplexer using anti-reflection contra-directional couplers for CWDM networks on silicon," Opt. Express, vol. 21, no. 6, pp. 6733-6738, 2013.
14. J. St-Yves, H. Bahrami, P. Jean, S. Larochelle, and W. Shi, "Widely bandwidth-tunable silicon filter with an unlimited free-spectral range," Opt. Lett., vol. 40, no. 23, pp. 5471-5474, 2015.
15. H. Qiu , J. Jiang, P. Yu , D. Mu, J. Yang, X. Jiang, H. Yu , R. Cheng, and L. Chrostowski, "Narrow-band add-drop filter based on phase-modulated grating-assisted contra-directional couplers," J. Lightwave Technol., vol. 36, no. 17, pp. 3760-3764, 2018.
16. D. Mu, H. Qiu, J. Jiang, X. Wang, Z. Fu, Y. Wang, X. Jiang, H. Yu, and J.Yang, "A four-channel DWDM tunable add/drop demultiplexer based on silicon waveguide Bragg gratings," Photon. J., vol. 11, no. 1, pp. 6600708, 2019.
17. K. Nakatsuhara, A. Kato, and Y. Hayama, "Latching operation in a tunable wavelength filter using Si sampled grating waveguide with ferroelectric liquid crystal cladding," Opt. Express, vol. 22, no. 8, pp. 9597-9603, 2014.

18. M. Caverley, X. Wang, K. Murray, N. A. F. Jaeger and L. Chrostowski, "Silicon-on-insulator modulators using a quarter-wave phase-shifted Bragg grating," IEEE Photon. Technol. Lett., vol. 27, no. 22, pp. 2331-2334, 2015.
19. H. Okayama, D. Shimura, Y. Onawa, H. Takahashi, S. Miyamura, H. Yaegashi and H. Sasaki, "Polarization conversion Si waveguide Bragg grating for polarization independent wavelength filter," Tech. Digest 11th Group IV Photonics, paper ThP19, Paris, 2014.
20. H. Okayama, D. Shimura, Y. Onawa, H. Yaegashi and H. Sasaki, "Polarization rotation Bragg grating using Si wire waveguide with non-vertical sidewall," Opt. Express, vol. 21, no. 25, pp. 31371-31378, 15 Dec, 2014.
21. H. Okayama, Y. Onawa, D. Shimura, H. Takahashi, S. Miyamura and H. Yaegashi, "Si wire waveguide polarisation-independent wavelength filter using polarisation rotation Bragg grating," Electron. Lett., vol. 50, pp. 1477-1478, 2014.
22. H. Yun, Z. Chen, Y. Wang, J. Fluekiger, M. Caverley, L. Chrostowski, and N. A. F. Jaeger, "Polarization-rotating, Bragg-grating filters on silicon-on-insulator strip waveguides using asymmetric periodic corner corrugations," Opt. Lett., vol. 40, no. 23, pp. 5578-5581, 2015.
23. H. Okayama, Y. Onawa, D. Shimura, H. Yaegashi and H. Sasaki, "Polarisation rotation Bragg grating with high diffraction efficiency using Si waveguide top surface groove grating," Electron. Lett., vol. 51, pp. 1909-1911, 2015.
24. H. Okayama, Y. Onawa, D. Shimura, H. Yaegashi and H. Sasaki, "Silicon waveguide polarization rotation Bragg grating with phase shift section and sampled grating scheme," Jpn. J. Appl. Phys., vol. 55, p. 082202, 2016.
25. H. Okayama, Y. Onawa, D. Shimura, H. Yaegashi and H. Sasaki, "Silicon waveguide polarization rotation Bragg grating with resonator cavity section," Jpn. J. Appl. Phys., vol. 56, no. 4, p.042502, 2017.
26. H. Okayama, Y. Onawa, D. Shimura, H. Yaegashi and H. Sasaki, "Silicon waveguide polarization rotator sampled Bragg grating," Opt. Lett., vol. 42, no. 11, pp. 2142-2144, 2017.
27. H. Yun, L. Chrostowski, and N. A. F. Jaeger, "Narrow-band, polarization-independent, transmission filter in a silicon-on-insulator strip waveguide," Opt. Lett., vol. 44, no. 4, pp. 847-850, 2019.
28. H. Okayama, Y. Onawa, D. Shimura, H. Sasaki, and H. Yaegashi, "Wavelength add/drop device using silicon waveguide polarization rotator grating," Tech. Digest Advanced Photonics Congress, paper Tu5A.3, Zurich, 2018.
29. H. Okayama, Y. Onawa, D. Shimura, H. Sasaki, and H. Yaegashi, "Asymmetric directional coupler type contra-directional polarization rotator Bragg grating: design," Jpn. J. Appl. Phys., vol. 58, p. 068002, 2019.
30. H. Okayama, Y. Onawa, H. Takahashi, D. Shimura, H. Yaegashi and H. Sasaki, "Silicon waveguide contradirectional coupler polarization rotation Bragg grating," Tech. Digest IEEE Photonics Conference, paper ThI3.3, 2020.
31. W. K. Burns and M. F. Milton, "Waveguide transitions and junctions," in Guided-wave optoelectronics, T. Tamir, Ed. (Springer-Verlag, 1988)
32. S.-H. Jeong, Y. Onawa, D. Shimura, H. Okayama, T. Aoki, H. Yaegashi, T. Horikawa, and T. Nakamura, "Polarization Diversified 16λ Demultiplexer Based on Silicon Wire Delayed Interferometers and Arrayed Waveguide Gratings," J. Lightwave Technol., vol. 38, pp. 2680-2687, 2020.